\documentclass[11pt]{article}
\usepackage[utf8]{inputenc}
\usepackage[T1]{fontenc}
\usepackage[margin=1.1in]{geometry}
\usepackage{amsmath,amssymb,booktabs,graphicx,setspace,natbib,caption}
\usepackage{xcolor}
\usepackage[colorlinks=true,citecolor=blue,linkcolor=blue]{hyperref}

\usepackage{float}
\begin{document}
\title{ \bfseries Spillover-Informed  Network Architecture for Global Volatility Forecasting}
\date{}
\maketitle
\vspace{-2cm}
\begin{center}
Neha Gupta$^{1}$,
Nishit Soni$^{2}$,
Aditya Maheshwari$^{3}$

\vspace{2mm}

\footnotesize
\begin{tabular}{l}

$^{1}$\emph{School of Computing (SoC), Dehradun Institute of Technology University, Dehradun, Uttarakhand, 248009, India.}\\
$^{\textrm{*}}$\emph{Corresponding author:neha.gupta@dituniversity.edu.in}\\

$^{2}$Indian Institute of Technology Ropar, Rupnagar 140001, Punjab, India.$^{}$ \emph{Email: soninishit256@gmail.com} \\

$^{3}$Operations Management and Quantitative Techniques Area, Indian Institute of Management Indore, 453556, India.\\
\emph{Email: adityam@iimidr.ac.in}
\end{tabular}

\end{center}


\begin{abstract}
\noindent
Spillover of volatility shocks across borders during turbulent periods makes accurate equity market volatility forecasts especially critical for risk management, derivatives pricing, and regulatory capital. In this paper, we  examine whether volatility forecasts improve when models incorporate information on how markets are connected, and whether the choice of connection measure matters. Using daily data on 29 equity indices from
every major region over 2015-2025, we let each market's forecast draw
on the recent volatility of the markets linked to it, with the strength
of each link set either by geography, return correlation, or the
Diebold-Yilmaz (DY) spillover network estimated from the data. The spillover-informed forecasting model achieved an approximately 13\% reduction in out-of-sample QLIKE loss relative to the standard Heterogeneous AutoRegressive benchmark ($p<0.001$) and attained the highest model confidence set $p$-value among the models considered. Several benchmark models that do not explicitly incorporate cross-market structure were excluded from the 90\% model confidence set.
The greatest improvements were observed during periods of elevated market stress: the COVID-19
crash, the Russian invasion of Ukraine, and the $2025$ US tariff shock,
spillover-informed forecasts achieved  approximately 21\% lower QLIKE loss than an otherwise identical network-blind model ($p = 0.018$), with no significant performance loss during calm periods. The advantage is economically material: a
volatility-targeting investor would pay $322$-$373$ basis points per year,
net of transaction costs, for spillover-informed forecasts, versus
an insignificant $85$-$102$ basis point for the alternatives. Finally,
when the model is left to infer connections on its own, it recovers
the DY network from the forecast objective alone
(permutation $p=0.0005$). 
\end{abstract}


\section{Introduction}

\noindent Shocks to one equity
market are transmitted to others within hours, carried by common
investors, shared fundamentals, and global risk sentiment, a
regularity documented since the ``meteor shower'' evidence of
\cite{engle1990} and the transmission studies of \cite{hamao1990},
and formalized for the modern era by the connectedness literature
\cite{diebold2009, diebold2012}. The COVID-19 crash erased roughly a
third of global equity value in five weeks across every time \cite{Baker2020,Ramelli2020}, and
the 2025 US tariff announcement moved markets from Tokyo to S\~ao Paulo
within a single trading day. Because next-day volatility is the direct
input to position sizing, option pricing, margin setting, and capital
provisioning \cite{poon2003}, and because forecast errors are most
expensive precisely when markets are turbulent and cross-border
linkages are strongest, the question of how to forecast volatility in
an interconnected system is of first-order practical importance.\\\\
\noindent The models in the realized-volatility forecasting literature, such as the Heterogeneous AutoRegressive (HAR) family
of \cite{corsi2009} and its extensions, treat each market as independent, conditioning tomorrow's variance on that market's own past alone. 
When a shock originates abroad and arrives overnight, a univariate model has no way to anticipate it.  It learns of the shock only after the local market has already absorbed it, and its forecast lags reality by a full day at the worst possible time.
The natural remedy is to allow each market's forecast to incorporate the recent volatility of other markets. 
With $N=29$ indices
there are over eight hundred possible directed linkages; estimating them
all freely invites overfitting, while the usual shortcuts (grouping by
region or by unconditional return correlation) encode assumptions about
the transmission channel that may simply be wrong. An exploration of an 
economically grounded answer to how volatility actually travels between
markets, and a test of whether using that answer improves forecasts is required.\\\\
\noindent The connectedness framework of \cite{diebold2009, diebold2012} provides  an answer. By decomposing the forecast-error variance of a vector autoregression, it measures how much of each market's
unpredictable movement is attributable to shocks originating in every
other market, yielding a directed, weighted network of spillovers
estimated entirely from data. This network is a standard descriptive
tool in the systemic-risk literature, but almost always as an
\emph{output}, used to describe the system after the fact. This paper investigates whether, it is useful as an \emph{input}: whether telling a forecasting
model that this is how markets are connected makes its forecasts better.\\\\
\noindent We investigate two questions in this paper. First, if volatility
propagates over a network of markets, do forecasts improve when the
forecasting model is informed about that network? Second, does the
choice of network matter, that is,  an economically measured map of spillovers
more valuable than the natural alternatives? We answer with a
specification in which each market's predicted next-day variance
depends not only on its own past but also on the past volatilities of
the markets connected to it, with the strength of each connection set
by one of three candidate networks, namely, geographic proximity, unconditional
return correlation, or the variance-decomposition spillover network of
\cite{diebold2012}. The forecasting model itself, described in
Section~\ref{section:4}, is drawn unchanged from the recent literature
on forecasting over networks and is held fixed throughout, so that any
difference in performance is attributable to the network supplied
rather than to the model. We evaluate one-day-ahead variance forecasts for $29$ equity
indices spanning every major region over 2020-2025, a period
containing the COVID-19 crash and recovery, the Russian invasion of
Ukraine, and the April $2025$ US tariff shock, precisely the conditions
under which cross-market linkages should matter most.\\\\
\noindent The evidence indicates that both questions can be answered  positively and can be summarized in four findings. First, conditioning on the
cross-section helps, that is,  every network-informed specification, and a
network-blind specification with the same information, improves on the
HAR benchmark by $12$-$13$\% in forecast loss, confirming that a market's
neighbors carry information its own history does not. Second, the
choice of network matters, and the economically grounded network
prevails. The Diebold-Yilmaz (DY) spillover network attains the top
model-confidence-set $p$-value ($1.00$); the $90\%$ set retains only
specifications that exploit the cross-section, while the correlation
network, every HAR variant (including a leverage-HAR), a
gradient-boosted benchmark, and the random walk are rejected. Third,  when the model is left to
infer the network from the data, using only the volatility-forecasting
objective, it recovers the DY structure (Mantel
permutation test, $p=0.0005$), that is,  two constructions with entirely
different foundations, a variance decomposition of returns and a purely
predictive criterion, converge on the same map of which markets move. Fourth, the advantage is economically significant and
concentrated where it should be. It is worth more than $320$ basis points per
year to a volatility-targeting investor net of transaction costs. Among all eleven specifications, only the network-informed models earn
gains statistically distinguishable from zero. Moreover, almost the entire
gain materializes in crisis windows, precisely when cross-border shocks cascade and accurate variance forecasts matter most.\\\\
The remainder of the paper is organized as follows.
Section~\ref{section:2} reviews the related literature.
Section~\ref{section:3} describes the data, the realized-variance
measure, and the event set. Section~\ref{section:4} develops the
connectedness measure, the candidate networks, and the forecasting
models. Section~\ref{section:5} presents the empirical results,
Section~\ref{section:6} the robustness checks and economic
interpretation, and Section~\ref{section:7} concludes.

\section{Related Literature}
\label{section:2}
This paper draws on three aspects of literature. The first is realized-volatility forecasting.  Intraday information based variance measures make volatility effectively observable \cite{andersen1998}, while the link between realized volatility and the conditional return distribution was formalized in subsequent work showing that simple time-series models for log realized variance provide accurate forecasts \cite{andersen2003}. The HAR model of \cite{corsi2009} remains the benchmark for the prediction of real volatility, with later studies improving the estimation by accounting for the measurement error in the real-variance proxies \cite{bollerslev2016}; \cite{poon2003} survey the broader forecasting literature and its practical uses.
On the evaluation, \cite{patton2011} showed that among common loss functions only QLIKE and MSE deliver noise-resistant rankings in the
volatility proxy, with QLIKE having superior power \cite{patton2009}.
We adopt this framework wholesale, with QLIKE is our headline loss, MSE on
log-variance is reported but de-emphasized, and all model comparisons
use \cite{diebold1995} tests with the \cite{harvey1997} correction
and the model confidence set of \cite{hansen2011}.

The second is the literature on cross-market linkages, contagion, and
connectedness. Early studies showed that volatility can spread from one financial market to another.
\cite{hamao1990} documented volatility spillovers across New York, London, and Tokyo markets, while \cite{engle1990} showed that foreign news affects domestic
volatility (``meteor showers'') over and above domestic persistence (``heat waves'').
Later, \cite{forbes2002} explained that higher volatility can exaggerate correlations, so much of the observed ``contagion'' reflects normal market interdependence. 
This finding motivated the use of measures beyond the simple correlation. The connectedness framework of \cite{diebold2009,
diebold2012, diebold2014} measures directional spillovers using the generalized forecast-error variance decomposition of \cite{pesaran1998}. Other network based approaches include those proposed by \cite{billio2012, barunik2018}.


The third is machine learning in financial forecasting, and in
particular the graph neural networks from which our model is drawn.
Modern graph convolutions trace to spectral constructions on graphs
\cite{defferrard2016} and their first-order simplification
\cite{kipf2017}, which the spatio-temporal literature then combined
with recurrent or convolutional dynamics for forecasting on networks, leading to models such as diffusion-convolutional RNNs \cite{li2018dcrnn}, Graph WaveNet
\cite{wu2019}, and adaptive-adjacency models \cite{bai2020} that learn
the graph end-to-end. These methods were developed largely for traffic
networks, where the graph is known. The financial setting is different because the underlying network must be measured or inferred from the data. Existing applications of deep learning to volatility
\cite[e.g.][]{bucci2020, gu2020} have mostly used fully-connected or
recurrent architectures without any economic spatial structure. 
Finally, our economic evaluation follows the
volatility timing tradition of \cite{fleming2001, fleming2003}, who
showed that variance forecasts have substantial value for a mean-variance investor.

\section{Data and Realized-Variance Construction}\label{section:3}

Evaluating spillover-informed forecasts imposes three requirements on the data that include a broad
cross-section of equity markets spanning all major regions and time zones, a daily measure of realized variance for each market, and a set
of well-identified systemic events around which forecast performance can
be examined. This section describes the set of equity indices and sample, the construction of returns and the range-based
realized-variance proxy, and the event set. The complete data cleaning, de-spiking, and missing-value rules are provided in \ref{appendix:A}.

\subsection{Sample}
We study $29$ headline equity indices covering North and Latin America, Europe, the Middle East, Africa, and Asia-Pacific, listed in Table~\ref{tab:universe}. Daily open, high, low, and close prices are collected from Yahoo Finance and span from $2$ January $2015$ to $30$ December $2025$, covering $2{,}866$ trading days on the union calendar, defined as the set of dates on which at least one index traded (see \ref{Appendix:A1}). The set of indices was selected before the forecasting analysis. To be included, an index was required to have data available for at least $50\%$ of the union-calendar days and to remain actively traded until the end of the sample period. Selecting the indices in advance, rather than after examining data availability, helps avoid survivorship-related look-ahead bias.\\
\noindent The selected indices represent a diverse set of global stock markets. They include large and highly liquid developed markets, such as those in North America and Western Europe, as well as smaller and more volatile emerging markets. The indices also cover different time zones, from the Asia-Pacific region to North America, allowing the study to capture volatility transmission across global markets. In addition, the markets exhibit a wide range of risk levels. Over the full sample period, the annualized volatility ranges from about $9\%$ for Singapore's STI to about $30\%$ for Argentina's MERVAL (Table~\ref{tab:universe}). This diversity ensures that the proposed models are evaluated under different market conditions rather than only on developed markets.
\begin{table}[!htb]
\centering

\begin{tabular}{llcc@{\qquad}llcc}
\toprule
Market & Region & Vol.\% & Market & Region & Vol.\% \\
\midrule
SP500 & N.~Am. & 12.8 & OMXS30 & Europe & 13.8 \\
DJIA & N.~Am. & 12.8 & SMI & Europe  & 12.0 \\
NASDAQ & N.~Am. & 15.4 & AEX & Europe  & 13.0 \\
TSX & N.~Am.  & 10.6 & BIST100 & Europe & 19.7 \\
IPC & L.~Am. & 13.7 & NIKKEI225 & Asia & 12.8 \\
BOVESPA & L.~Am.  & 19.1 & HSI & Asia  & 15.4 \\
MERVAL & L.~Am. & 30.1 & SSE & Asia  & 15.9 \\
FTSE100 & Europe & 12.7 & KOSPI & Asia & 12.0 \\
DAX & Europe & 14.4 & TWII & Asia & 10.8 \\
CAC40 & Europe  & 14.0 & SENSEX & Asia  & 12.7 \\
IBEX35 & Europe  & 15.0 & STI & Asia & 9.4 \\
BEL20 & Europe & 12.9 & JKSE & Asia &  12.1 \\
ATX & Europe &  17.0 & ASX200 & Pacific &  10.9 \\
TA125 & Mid.~East  & 11.3 & JSE & Africa &  13.6 \\
TASI & Mid.~East  & 12.1 & & & & \\
\bottomrule
\end{tabular}
\caption{The $29$ market indices used in this study. Vol.\% denotes the full-sample annualized volatility of each market index.}
\label{tab:universe}
\end{table}

\subsection{Returns and Realized-Variance Construction}
We compute log returns as differences in local-currency closing prices and scale them by $100$. Daily realized variance is proxied using the range-based estimator of \cite{garman1980},
\begin{equation}
\mathrm{GK}_{i,t} \;=\; \tfrac{1}{2}\bigl(\ln H_{i,t}/L_{i,t}\bigr)^2
\;-\;\bigl(2\ln 2 - 1\bigr)\bigl(\ln C_{i,t}/O_{i,t}\bigr)^2,
\label{eq:gk}
\end{equation}
scaled to percentage-squared units. The range estimator exploits the
full intraday path summarised by the open, high, low, and close, and is
therefore far more informative about a day's variance than the squared
close-to-close return. Under standard assumptions, it is five to eight
times more efficient, making it possible to construct a daily realized-variance panel across $29$ markets without intraday transaction data.
Because the estimator uses only within-session prices, it excludes the overnight return. As a result, markets with large overnight gaps may have
modestly understated variance levels. However, this affects only the variance levels and leaves the QLIKE-based ranking of forecasts unchanged (see~\ref{Appendix:A2}).\\\\
\noindent Three data-cleaning steps are applied to reduce errors in the raw data. First, unusually large one-day returns caused by data errors are removed. Second, extreme return values in two markets are winsorized. Third, days with a zero price range are excluded. The missing values created by these steps, together with those caused by different market holidays (about $5.70\%$ of the data), are filled using the median value only for model input. These imputed values are not used when evaluating the forecasts. The complete cleaning procedure, thresholds, and imputation rules are provided in \ref{Appendix:A3}. The models predict the natural logarithm of the Garman--Klass volatility measure, $y_{i,t}=\ln \mathrm{GK}_{i,t}$, which is approximately normally distributed. Since the true daily volatility is unobservable, forecast performance is evaluated using the QLIKE loss, which provides reliable model comparisons even when volatility is measured using a proxy \cite{patton2011}.

\subsection{The Event Set}

Seven systemic events are used for the descriptive analysis and the
crisis based forcasting evaluation. These events include the Chinese equity crash ($24$ Aug $2015$),
the Brexit referendum ($24$ Jun $2016$), the Trump election rally ($9$ Nov
$2016$), the COVID-$19$ crash ($12$ Mar $2020$), the COVID recovery rally
($24$ Mar $2020$), the Russian invasion of Ukraine ($24$ Feb $2022$), and the
US ``Liberation Day'' tariff announcement ($3$ Apr $2025$). The events were
chosen ex ante as discrete, major global shocks with clearly defined dates, spread across the sample so that both the
training and out-of-sample windows include important market stress events. The first three events fall within the training sample and are used only in the connectedness analysis of Section~\ref{section:5}. The remaining four events occur in the out-of-sample period and are used only to evaluate forecast performance during crisis periods.


\section{Empirical Framework}\label{section:4}

\subsection{Measuring Connectedness with the Diebold-Yilmaz Spillover Network}\label{section:4.1}

We need a measure of how strongly each market is connected to each other
market that is (i) directional, since the influence of the US on Japan
need not equal the reverse, (ii) estimated from data rather than
assumed, and (iii) economically interpretable. The connectedness framework of \cite{diebold2009, diebold2012} provides exactly this. Its core idea is simple. Fit a vector auto-regression to the markets' returns and
ask of the unpredictable part of market $i$'s return over the next $H$
days (its forecast-error variance), how much can be traced to
shocks that originated in market $j$? If a large share of what
surprises market $i$ comes from market $j$, then $j$ is strongly
connected to $i$. If none does, the two markets are considered unconnected. Computing this
share for every ordered pair $(i,j)$ produces a complete directed
network of who-moves-whom, with each link's weight measured in the
natural units of forecast-error variance and the whole network read off
from the estimated dynamics rather than imposed by the researcher. This
is why we adopt it as our economic network. It provides the framework in the literature for answering the question, how a
shock in one market propagate to the others?

\noindent Formally, let $r_t$ be the $29\times 1$ return vector. We estimate a
VAR($p$) with $p=5$ lags and ridge regularization (required for
stability with $29$ variables and $5$ lags, which would otherwise demand
estimating over four thousand auto-regressive coefficients) and compute
the $H$-step generalized forecast-error variance
decomposition of \cite{pesaran1998} at horizon $H=10$ days, which,
unlike the older
Cholesky decomposition, does not depend on the arbitrary ordering of
the markets. The quantity $\theta_{ij}(H)$ is the share of
market $i$'s $H$-step forecast-error variance attributable to shocks in
market $j$. Row-normalizing $\tilde\theta_{ij} =
\theta_{ij}/\sum_k \theta_{ik}$ gives the DY connectedness table. The
off-diagonal grand mean is total connectedness, and the network we
supply to the model is the off-diagonal matrix $A^{\mathrm{dy}}_{ij}=\tilde\theta_{ij}$,
$i\neq j$. On the full training sample, total connectedness is $80.2\%$, which is four-fifths of forecast-error variance in this system is of
foreign origin, and the rolling $252$-day version (Figure~\ref{fig:roll})
spikes above $90\%$ in the COVID period and jumps visibly at the $2025$
tariff announcement. In-sample estimation of $A^{\mathrm{dy}}$ uses
only data through the end of the initial training window. The matrix is re-estimated every six months using all data available up to that time. This ensures that no future information is used to make forecasts.

\subsection{Candidate Networks and Normalization}

We compare three fixed networks. The first is \emph{geographic}
($A^{\mathrm{geo}}$: weight $1$ for same-region pairs, $0.5$ for
adjacent-session pairs). The second is \emph{correlation} ($A^{\mathrm{corr}}$:
$k$-nearest-neighbour graph on training-sample return correlations,
$k=5$). The third is the DY network described above. All graphs are row-normalized with a
guaranteed self-weight,
\begin{equation}
\hat A_{ii} = \tfrac12, \qquad
\hat A_{ij} = \tfrac12\,\frac{A_{ij}}{\sum_{k\neq i} A_{ik}},\;\; j\neq i,
\label{eq:norm}
\end{equation}
so that a market's own signal always carries half the weight of the
graph convolution regardless of neighbourhood size. This is important because, under
the conventional $D^{-1}(A+I)$ normalization, a market inside the dense
ten-member European block retains only a $10\%$ self-weight, and the
network-weighted average collapses toward a regional mean, washing out
market-level information. Equation~\eqref{eq:norm} removes this effect for all graphs symmetrically.

\subsection{Benchmark Models and Network-Based Forecasting Approaches}
 We compare the proposed model with several benchmark methods. The Random Walk (RW) model uses the current value as the forecast, i.e., $\hat y_{i,t+1}=y_{i,t}$. The HAR model predicts future volatility using the daily, weekly ($5$-day), and monthly ($22$-day) averages of $y_{i,t}$, with the model estimated separately for each market using an expanding window. HARX extends HAR by adding the average volatility across all markets as a simple measure of global market conditions. This allows us to examine whether any improvement comes only from using information from other markets.\\
\noindent We also consider two stronger benchmark models. LHAR extends HAR by including daily, weekly, and monthly aggregated negative returns to capture the leverage effect \cite{corsi2012}. It is the most advanced HAR-based model that can be applied using only daily OHLC data, since models such as HARQ and semivariance HAR require intraday measures that are not available in our dataset \cite{bollerslev2016}. Finally, we include a Gradient Boosting Machine (GBM) \cite{friedman2001}, a widely used machine learning model. GBM is trained using the same input variables as the proposed model, including HAR features, leverage variables, the global volatility factor, and a market indicator, under the same expanding-window framework. This benchmark tests whether a flexible machine learning model can capture cross-market relationships without explicitly modeling the market network.\\
\noindent Next, we include the Network-Blind Model as a flexible benchmark. We use a Long Short-Term Memory (LSTM) network \cite{hochreiter1997}, which uses the past $L=22$ days of volatility data from all markets to jointly predict the next-day volatility of each market. Unlike the network-based models, the LSTM is not provided with any information about market connections and must learn cross-market relationships directly from the data. Comparing its performance with network-based models helps measure the benefit of explicitly including market relationships.\\
\noindent Furthermore, we consider the Network-Based Model that uses the information from the market connections. Since markets are linked with each other, the model should be able to capture both the relationships across different markets and the changes in each market's volatility over time. A graph-convolutional recurrent network satisfies both requirements and the graph-convolutional layer combines information from connected markets. Specifically, the operation $\hat{A}Z$ replaces each market's feature vector with a weighted average of the feature vectors of its neighbouring markets, where the weights are determined by the DY network. As a result, the forecast for each market uses information from the markets that are most strongly connected to it. The combined information is then transformed using a coefficient matrix $W$, whose values are learned from the data. The temporal component of the model is a Gated Recurrent Unit (GRU), which captures how volatility evolves over time. The GRU maintains a hidden state that carries useful information from previous days and updates it as new data become available. Concretely, the graph-convolutional GRU (GCGRU) of \cite{seo2018} replaces the
input and recurrent transforms of an ordinary GRU with the
network-weighted operation $Z \mapsto \hat A Z W$:
\begin{align}
u_t &= \sigma\!\bigl(\hat A\,[x_t, h_{t-1}]\,W_u\bigr), \quad
r_t = \sigma\!\bigl(\hat A\,[x_t, h_{t-1}]\,W_r\bigr), \nonumber\\
\tilde h_t &= \tanh\!\bigl(\hat A\,[x_t, r_t \odot h_{t-1}]\,W_h\bigr), \quad
h_t = u_t \odot h_{t-1} + (1-u_t)\odot \tilde h_t,
\label{eq:gcgru}
\end{align}
where $x_t \in \mathbb{R}^{29\times d}$ contains lagged log-variance
features per market and $\hat A$ is one of the three fixed networks.
\begin{figure}[H]
    \centering
   \includegraphics[width=0.9\textwidth]{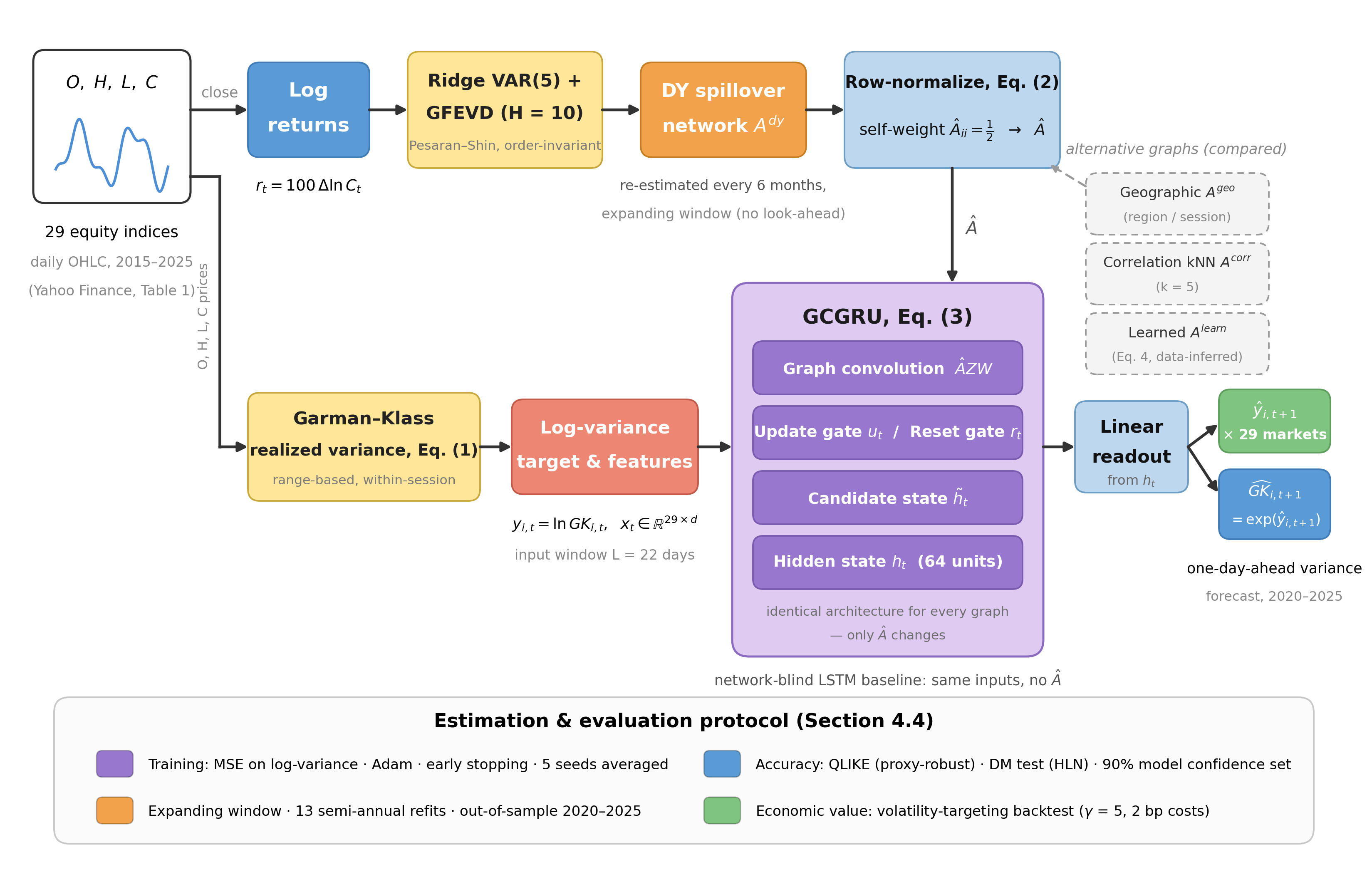}
    \caption{The spillover-informed forecasting pipeline. Log returns yield the Diebold--Yilmaz network~\ref{section:4.1}, while OHLC prices yield the Garman--Klass log-variance inputs (Eq.~\ref{eq:gk}). The normalized graph $\hat{A}$ (Eq.~\ref{eq:norm}) enters the GCGRU (Eq.~\ref{eq:gcgru}), which outputs next-day log-variance forecasts for all $29$ markets.}
    \label{fig:workflow}
\end{figure}
\noindent Moreover, we also consider a model that learns the market network directly from the data instead of using a predefined network. AdaptGCGRU is an adaptive graph-convolutional recurrent model that learns the market relationships directly from the data during training. Following \cite{bai2020}, the model learns the connections through two sets of trainable market embeddings, $E_1, E_2 \in \mathbb{R}^{29\times e}$, and constructs the network as
\begin{equation}
A^{\mathrm{learn}} = \mathrm{softmax}\bigl(\mathrm{ReLU}(E_1 E_2^\top)\bigr),
\label{eq:adapt}
\end{equation}
where the learned network is estimated jointly with the forecasting objective. This allows the model to identify market connections that are useful for volatility forecasting. We also consider a second variant, AdaptGCGRU\_dy, which combines the learned network with the DY network:
$$
A^{\mathrm{eff}} = (1-g)A^{\mathrm{learn}} + g\,\hat A^{\mathrm{dy}},
$$
where $g=\sigma(\gamma_0)$ is a learnable weight. The learned network is updated every six months using the available information up to that point and remains fixed between updates.\\
\noindent We use the following abbreviations for the graph-based models. GCGRU\_dy denotes the GCGRU model using the DY spillover network, GCGRU\_geo uses the geographic network, and GCGRU\_corr uses the correlation-based network. AdaptGCGRU denotes the adaptive GCGRU that learns the graph from the data, while AdaptGCGRU\_dy denotes the adaptive GCGRU initialized with the DY network. The LSTM model serves as the network-blind baseline.


\subsection{Estimation Procedure and Forecast Evaluation}
The out-of-sample period covers the period from $2$ January $2020$ to $30$ December $2025$. All models are re-estimated approximately every six months using an expanding window approach. This results in $13$ model updates during the evaluation period. At each update, the models are trained using all available data from the beginning of the sample up to that date, ensuring that no future information is used in the forecasting process. The network-based and network-blind models are
estimated by minimizing mean squared error (MSE) on log-variance using the
Adam optimizer (learning rate $10^{-3}$, batch size $64$).The models are trained for a maximum of $100$ epochs on each estimation window. The last $15\%$ of the data in each estimation window is used as a validation set. Training is stopped early if the validation loss does not improve for $10$ consecutive epochs, and the model parameters with the lowest validation loss are selected. All neural models
use identical settings, a hidden state size of $64$, an input window of $L=22$ trading days, and, for the data-inferred network models, a market embedding dimension of $e=16$. Keeping these settings identical ensures that differences in performance are due to the network structure rather than model tuning. Each model is trained using five different random initializations. The reported forecasts are the average across these five runs, and we also report the worst-performing run to ensure that the results are not driven by a favourable initialization.
Since the true daily variance is not directly observable and we use the Garman--Klass measure as a proxy, the choice of loss function is important. An inappropriate loss function may lead to incorrect comparisons between forecasting models because of the noise present in the volatility proxy. Patton \cite{patton2011} shows that the QLIKE loss provides consistent forecast rankings even when the variance is measured with error. It also penalizes volatility under-predictions more strongly, which is important for risk management. Therefore, we use QLIKE as our main evaluation metric
\begin{equation}
\mathrm{QLIKE}_{i,t} = \frac{\mathrm{GK}_{i,t}}{\widehat{\mathrm{GK}}_{i,t}}
- \ln\frac{\mathrm{GK}_{i,t}}{\widehat{\mathrm{GK}}_{i,t}} - 1,
\qquad \widehat{\mathrm{GK}}_{i,t} = \exp(\hat y_{i,t}),
\label{eq:qlike}
\end{equation}
which is zero for a perfect forecast and positive otherwise. To test whether one model performs significantly better than another, we use the Diebold--Mariano test \cite{diebold1995} with the small-sample correction of \cite{harvey1997}. The test is applied to the daily average loss differences across all markets, using HAR as the benchmark model. A pairwise test only compares two models at a time and cannot identify the best group of models when many models are considered. Therefore, we use the model confidence set approach of \cite{hansen2011}, which removes models that perform significantly worse and provides a set of models that contains the best model with $90\%$ confidence. We implement the model confidence set using the range ($T_{\max}$) statistic and a moving-block bootstrap with a block length of $10$ trading days and $1{,}000$ replications to account for dependence in daily loss differences. We also evaluate the economic value of forecasts using a volatility-targeting strategy. Each market is assigned a weight $w_{i,t}=\min(\tau/\hat\sigma_{i,t},\,4)$ with a daily volatility target of $\tau=1\%$. The portfolio is equally weighted across markets, includes a transaction cost of $2$ basis points, and is evaluated using certainty-equivalent returns based on mean-variance utility with risk aversion parameter $\gamma=5$ \cite{fleming2001}.

\section{Empirical Results}\label{section:5}

\subsection{The Dynamics of Connectedness}\label{section:5.1}

\begin{figure}[H]
\centering
\includegraphics[width=1.05\textwidth]{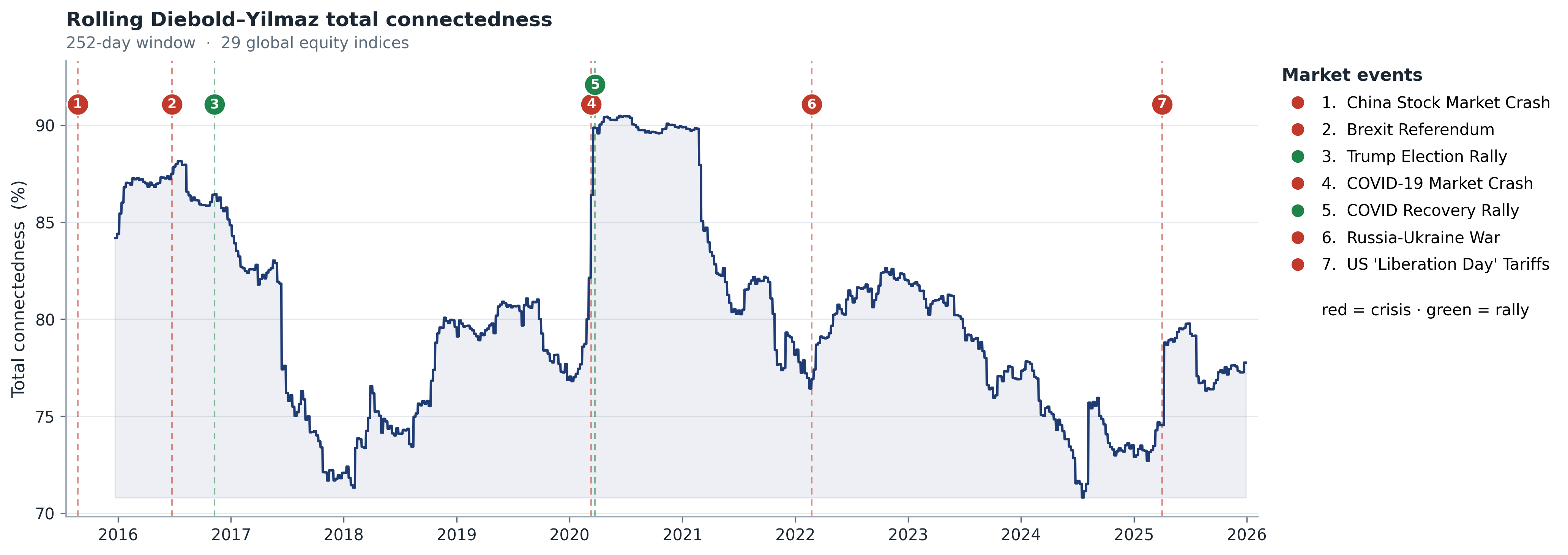}
\caption{Rolling DY total connectedness ($252$-day window,
$H=10$ GFEVD). Dashed lines mark the seven events;
the first three lie inside the initial training sample.}
\label{fig:roll}
\end{figure}

Figure~\ref{fig:roll} plots rolling $252$-day total connectedness among the
$29$ equity indices. Three observation are important. First, connectedness
remains high throughout the sample period (around 71-90\%), indicating that global equity markets are strongly linked.  
This provides motivation for using
models that incorporate cross-market information.\\
Second, connectedness
increases during major crisis periods. It rises above $90\%$ during the
COVID-$19$ market crash and remains elevated for several months, with another
increase observed around the $2025$ tariff announcement. This suggests that the
importance of market spillovers may vary across different market conditions, which is examined in Section~\ref{section:5.4}.

\begin{figure}[H]
\centering
\includegraphics[width=0.95\textwidth]{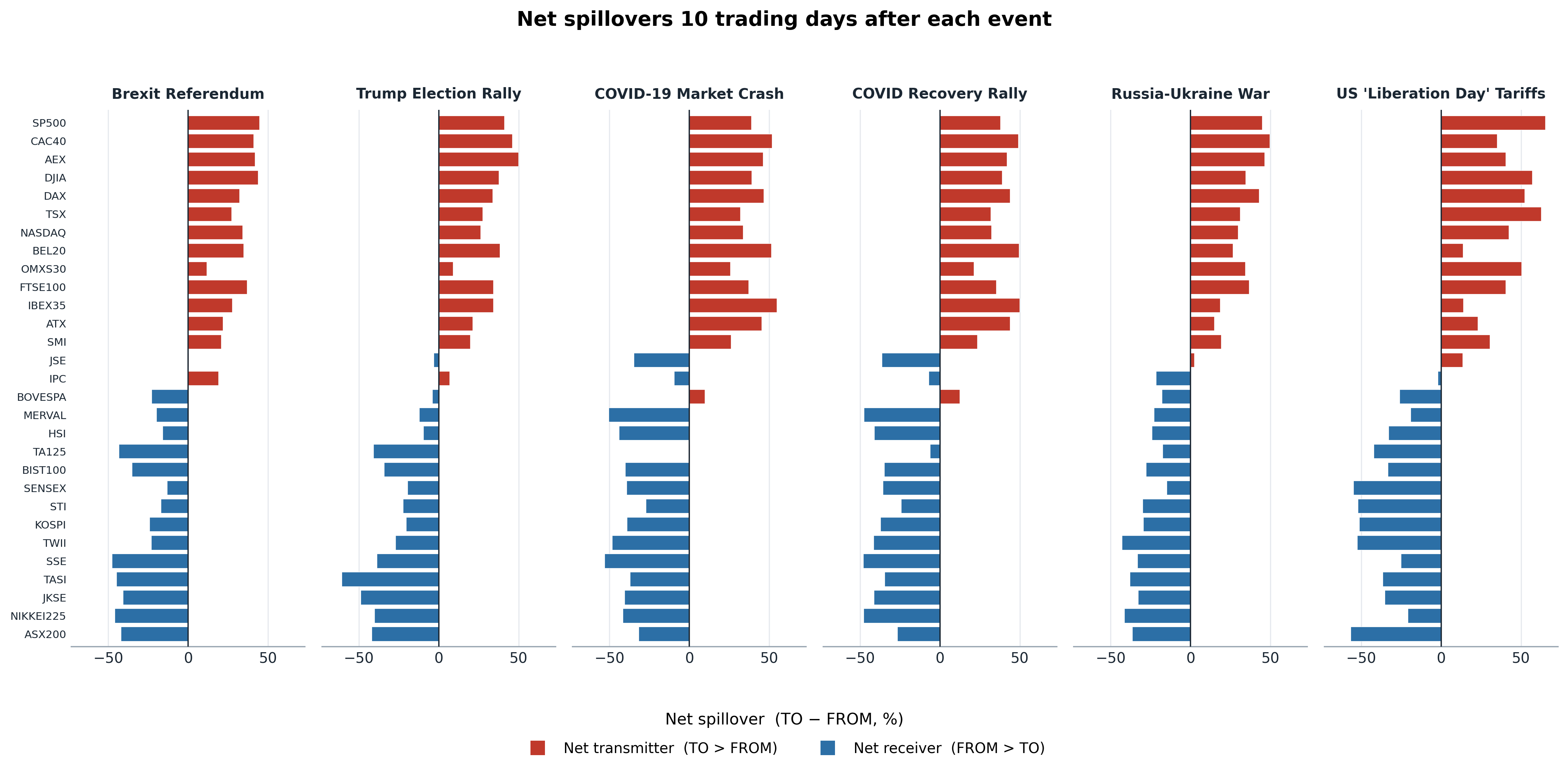}
\caption{Net directional spillovers (TO - FROM, \%) over the $10$ trading days after each event with an available post-event window. Red bars indicate net transmitters and blue bars indicate net receivers. Directional patterns may partly reflect time-zone effects and are interpreted qualitatively.}
\label{fig:net}
\end{figure}

\noindent Third, the directional spillover patterns in Figure~\ref{fig:net} are broadly
consistent across events with a 10-day post-event window. US markets, TSX,
DAX, and other major European markets generally act as net transmitters,
whereas many Asia-Pacific markets behave as net receivers. However, these
directional results may partly reflect differences in trading hours, as markets
that close later can mechanically receive information from earlier markets.
Therefore, we interpret the direction of spillovers qualitatively.

\subsection{Main Forecasting Results}
The connectedness analysis shows that the $29$ markets are strongly linked and that these connections become stronger during crisis periods. We next examine whether forecasting models can improve by using this market information. Table~\ref{tab:main} presents the out-of-sample one-day-ahead volatility forecast results for all eleven models during $2020$--$2025$. The models are ranked according to QLIKE loss, along with Diebold--Mariano test results against the HAR benchmark and model confidence set $p$-values.

\begin{table}[H]
\centering\footnotesize

\begin{tabular}{lccccc}
\toprule
Model & QLIKE & MSE$_{\log RV}$ & DM vs HAR & $p$ & MCS $p$ \\
\midrule
GCGRU\_dy        & \textbf{0.3430} & 0.5583 & $+5.05$ & $<0.001$ & \textbf{1.000} \\
AdaptGCGRU       & 0.3452 & 0.5556 & $+4.99$ & $<0.001$ & \textbf{0.563} \\
AdaptGCGRU\_dy   & 0.3453 & 0.5607 & $+4.77$ & $<0.001$ & \textbf{0.396} \\
GCGRU\_geo       & 0.3456 & 0.5604 & $+4.88$ & $<0.001$ & \textbf{0.180} \\
LSTM             & 0.3482 & 0.5608 & $+5.60$ & $<0.001$ & \textbf{0.563} \\
GCGRU\_corr      & 0.3485 & 0.5626 & $+4.86$ & $<0.001$ & 0.035 \\
LHAR             & 0.3605 & 0.5729 & $+4.06$ & $<0.001$ & 0.006 \\
GBM              & 0.3635 & 0.5740 & $+3.81$ & $<0.001$ & 0.042 \\
HARX             & 0.3641 & 0.5787 & $+4.82$ & $<0.001$ & 0.015 \\
HAR              & 0.3956 & 0.6007 & -       & -        & 0.022 \\
RW               & 0.6075 & 0.8902 & $-11.67$ & $<0.001$ & 0.000 \\
\bottomrule
\end{tabular}
\caption{Out-of-sample one-day-ahead forecast accuracy, 2020-2025
(29 markets $\times$ 1{,}565 days). DM is the Diebold-Mariano statistic
on daily cross-market QLIKE differentials versus HAR
(positive = better than HAR), with HLN small-sample correction.
MCS $p$ is the model-confidence-set $p$-value; models with $p>0.10$
(bold) survive in the $90\%$ set. The network-based and network-blind models are estimated from five random starting values and averaged.}
\label{tab:main}
\end{table}
\noindent Table~\ref{tab:main} highlights four main findings. First, using the full cross-section improves forecasting. All network-based models outperform HAR by $12$--$13$\% in QLIKE with DM statistics close to $5$, and the network-blind model also performs better than HAR. Second, stronger classical benchmarks reduce the gap but do not eliminate it. The leverage-HAR ($0.3605$) and gradient-boosted tree ($0.3635$) recover about two-thirds of HAR's improvement, but the cross-sectional neural models still provide the best performance.\\
Third, GCGRU\_dy ranks first with a QLIKE of $0.3430$ and achieves the highest MCS $p$-value ($1.000$). The 90\% model confidence set retains the DY-informed, adaptive, and geographic graph specifications together with the network-blind LSTM, whereas the correlation-network specification and the classical benchmarks are excluded. Fourth, GCGRU\_dy consistently outperforms the network-blind LSTM, and, as shown in Section~\ref{section:5.4}, this advantage is most pronounced during crisis periods when cross-market spillovers are strongest.\\
The DY network outperforms the correlation and geographic networks because it measures directional volatility transmission rather than simple co-movement or geographic proximity. Consequently, the graph captures economically meaningful channels through which volatility propagates across markets.



\subsection{The Learned Network Recovers the Spillover Structure}
\begin{figure}[t]
\centering
\includegraphics[width=\textwidth]{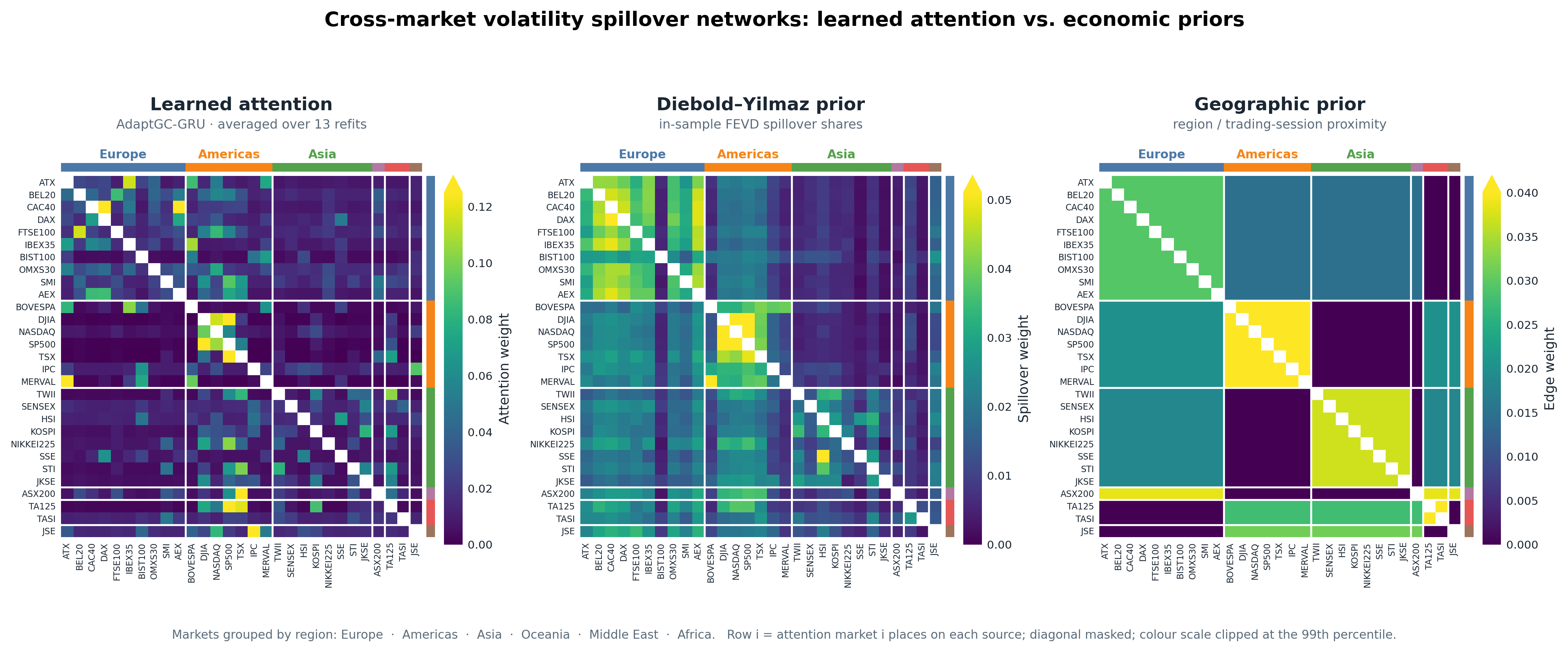}
\caption{Comparison of the data-inferred attention network (averaged over 13 semi-annual refits), the DY network, and the geographic network. Markets are grouped by region (Europe, Americas, Asia, Oceania, Middle East, and Africa). Row $i$ shows the weights assigned by market i to all source markets. Diagonals are masked, and the colour scale is clipped at the $99$th percentile. Additional diagnostics of the learned network are reported in \ref{appendix:B}.}
\label{fig:adj}
\end{figure}

Figure~\ref{fig:adj} compares the network learned by the data-inferred model with the DY and geographic networks. The learned network captures the main features of the DY network, including the dense US block (DJIA, NASDAQ, SP500, TSX), the European core, and the stronger connections from the US to Asia, even though no prior network information is provided. The agreement is statistically significant. We compare the learned network, averaged over the 13 refits, with the DY network using only the off-diagonal entries. The Spearman correlation between the two is $\rho = 0.254$. A Mantel permutation test with 2{,}000 permutations rejects the null of no association ($p = 0.0005$): none of the permuted networks produced a correlation as high as the observed one. This represents one of the principal findings of the study. A model provided with no explicit graph information learns an attention structure that exhibits a statistically significant correspondence with the network identified independently by the Diebold–Yilmaz (DY) methodology. This comparison is based exclusively on the graph learned by the data-driven model, as the blended variant incorporates the DY network as prior information and would therefore exhibit a mechanically induced correlation ($\rho = 0.70$), making it unsuitable for assessing network recovery.
The recovered structure is also interpretable at the market level. The learned graph identifies DJIA, NASDAQ, and TSX as the three strongest neighbours of SP500, while CAC40 is most strongly connected to AEX, DAX, and IBEX35. These regional groupings are consistent with those identified by the DY decomposition.
Together with the model confidence set results, these findings suggest that forecast-error spillovers are the main channel of volatility transmission, and a flexible model can learn this structure without prior network information. That the gated variant
AdaptGCGRU\_dy performs identically to its parents (QLIKE $0.3453$)
indicates the learned and DY graphs carry largely redundant information, as they should if both measure the same object.

\begin{figure}[H]
\centering
\includegraphics[width=0.62\textwidth]{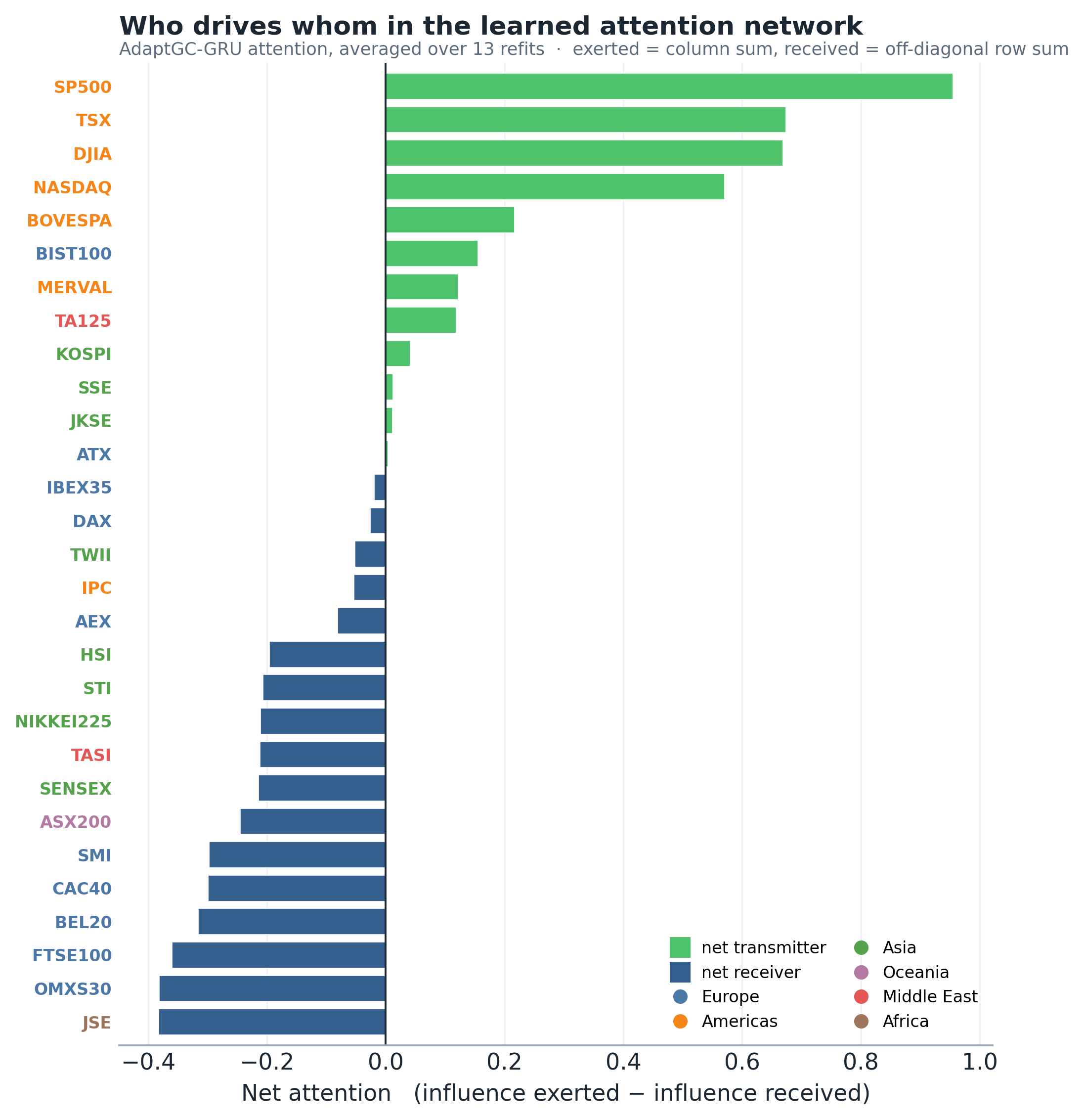}
\caption{Net attention in the learned network (influence exerted minus
influence received, averaged over the 13 refits). Label colours denote
regions.}
\label{fig:netattn}
\end{figure}

\noindent The inferred network's directional content can be summarized in one
statistic per market: its net influence, the total weight the other
markets place on it minus the total weight it places on them (labelled
``net attention'' in the figure, following the terminology of the
estimation method).
Figure~\ref{fig:netattn} ranks the markets by their net transmission strength. The four strongest transmitters in the learned graph are SP500, TSX, DJIA, and NASDAQ, matching the North American markets identified by the DY z network as the main sources of volatility spillovers (Figure~\ref{fig:net}). In contrast, the smaller European markets and the Asia-Pacific markets are the strongest receivers. This shows that the forecasting model independently recovers the same overall transmission pattern.

\subsection{Crisis-Conditional Performance}\label{section:5.4}

Table~\ref{tab:regime} divides the out-of-sample period into 71 crisis days, defined as the 20 trading days following each of the four out-of-sample events, and the remaining calm days. We focus on AdaptGCGRU\_dy because it includes both the learned and DY graphs. The results for GCGRU\_dy are very similar, showing that both models behave almost identically.
\begin{table}[H]
\centering\small

\begin{tabular}{llcccc}
\toprule
Benchmark & Regime & QLIKE bench & QLIKE AdaptGCGRU\_dy & DM & $p$ \\
\midrule
HAR  & crisis & 1.009 & 0.475 & $+2.75$ & 0.008 \\
HAR  & calm   & 0.367 & 0.339 & $+4.96$ & $<0.001$ \\
LSTM & crisis & 0.603 & 0.475 & $+2.42$ & 0.018 \\
LSTM & calm   & 0.336 & 0.339 & $-0.99$ & 0.323 \\
GCGRU\_dy & crisis & 0.486 & 0.475 & $+1.55$ & 0.126 \\
GCGRU\_dy & calm   & 0.336 & 0.339 & $-2.35$ & 0.019 \\
\bottomrule
\end{tabular}
\caption{QLIKE by regime. Crisis = 71 days in $[0,+20]$ windows after
the four out-of-sample events; calm = all other days. DM tests the
daily loss differential of AdaptGCGRU\_dy against each benchmark
within the regime.}
\label{tab:regime}
\end{table}

\noindent This regime analysis highlights the paper's main economic result. The DY-network model outperforms HAR in both calm and crisis periods, with much larger gains during crises. During crisis windows, it reduces QLIKE by 53\%, showing that cross-market information becomes most valuable when forecasts matter most. Compared with the network-blind model, there is no meaningful difference in calm periods, but the DY-network model achieves a 21\% lower QLIKE during crises ($0.475$ vs.\ $0.603$, $p=0.018$). \\
\begin{figure}[!htb]
\centering
\includegraphics[width=0.7\textwidth]{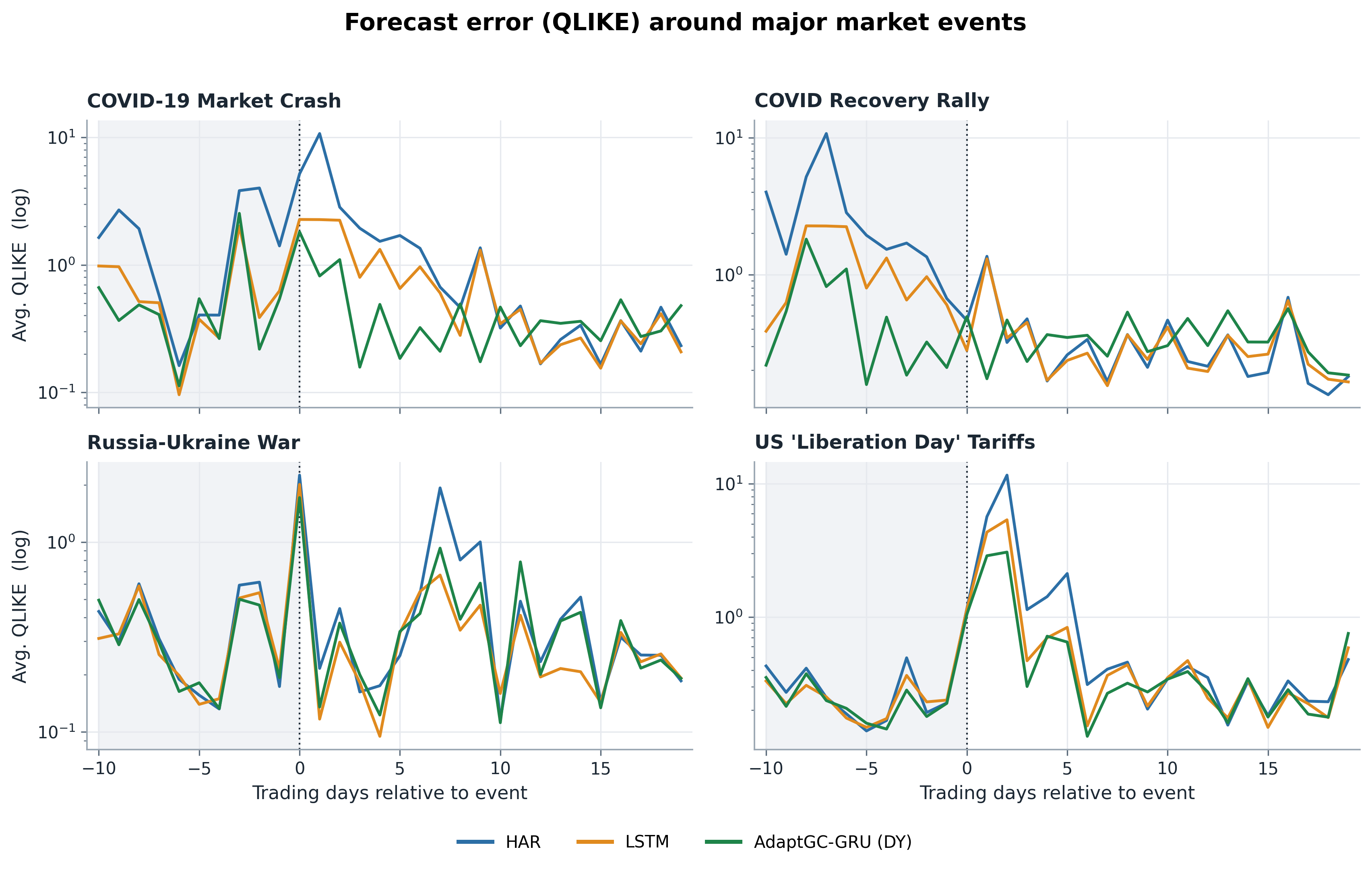}
\caption{Average QLIKE across 29 markets in event time (log scale),
for the four out-of-sample events. Day 0 is the event date.}
\label{fig:events}
\end{figure}

Figure~\ref{fig:events} traces this day by day around each
event. The ordering HAR $\succ$ network-blind $\succ$ DY-network in
post-event loss is visible in all four episodes. The two DY rows show
the fixed-network and blended models are statistically interchangeable
in crises, with a small calm-period edge to the fixed-network model.

\subsection{Economic Performance}

Table~\ref{tab:econ} converts forecasts into a volatility-targeting
strategy with $2$ bp turnover costs.

\begin{table}[H]
\centering\small

\begin{tabular}{lcccccc}
\toprule
Model & Ann.\ ret.\,\% & Sharpe$_{\text{net}}$ & MaxDD\,\% & CE\,\% & CE gain (bp) & $p$ \\
\midrule
AdaptGCGRU\_dy  & 14.9 & 1.08 & $-26.3$ & 10.1 & 373 & $<0.001$ \\
AdaptGCGRU      & 14.8 & 1.06 & $-27.4$ & 9.9  & 353 & $<0.001$ \\
GCGRU\_dy       & 14.4 & 1.04 & $-25.9$ & 9.6  & 322 & $<0.001$ \\
GCGRU\_geo      & 13.8 & 0.99 & $-26.5$ & 8.9  & 254 & 0.003 \\
GCGRU\_corr     & 13.0 & 0.93 & $-27.2$ & 8.1  & 173 & 0.026 \\
HARX            & 12.9 & 0.92 & $-27.9$ & 8.0  & 160 & 0.003 \\
LHAR            & 12.5 & 0.88 & $-27.1$ & 7.4  & 102 & 0.189 \\
LSTM            & 12.2 & 0.87 & $-28.3$ & 7.3  & 88  & 0.189 \\
GBM             & 12.3 & 0.87 & $-28.5$ & 7.2  & 85  & 0.253 \\
HAR             & 11.8 & 0.80 & $-31.2$ & 6.4  & 0   & - \\
RW              & 11.0 & 0.76 & $-27.8$ & 5.8  & $-57$ & 0.663 \\
\bottomrule
\end{tabular}
\caption{Volatility-targeting backtest, 2020-2025. The strategy applies a daily volatility target of 1\%, a leverage cap of 4, and equal weights across the 29
markets, with proportional transaction costs of 2\,bp per unit of turnover. All
reported statistics, including annualized return, net Sharpe ratio, maximum
drawdown, and certainty equivalent (CE), are net of these costs. CE gain is the
annualized certainty-equivalent gain over HAR under mean-variance utility ($\gamma = 5$); $p$ is from a Diebold-Mariano test on the daily utility differentials against HAR.}
\label{tab:econ}
\end{table}
 \noindent The DY-informed models deliver the highest economic value, with certainty-equivalent (CE) gains of $322$-$373$ basis points (bp) over HAR, higher net Sharpe ratios ($1.04$-$1.08$ vs.\ $0.80$), and lower maximum drawdowns. In contrast, the gains of the network-blind LSTM ($88$ bp), LHAR ($102$ bp), and GBM ($85$ bp) are not statistically significant, showing that the economic benefits mainly come from incorporating cross-market network information.\\
 \noindent The economic ranking is consistent with the forecasting results. The DY-informed models perform best, followed by the geographic and correlation network models. HARX provides a smaller but significant improvement, whereas LHAR, the network-blind LSTM, and GBM do not provide significant gains. The random walk performs worst because its high turnover leads to larger transaction costs.


\section{Robustness and Interpretation}\label{section:6}

\subsection{Component Analysis}
Table~\ref{tab:ablation} examines the contribution of each component by removing or replacing one part of the full model (learned attention with the DY prior) at a time. All models are trained using the same protocol, and Figure~\ref{fig:ablfig} in \ref{appendix:B} summarizes the results.\\
\begin{table}[H]
\centering\small
\begin{tabular}{llcccc}
\toprule
Model & Component removed / changed & QLIKE & $\Delta$\,\% & Crisis & $\Delta_{\mathrm{cr}}$\,\% \\
\midrule
AdaptGCGRU\_dy & none (full model: learned $\oplus$ DY prior) & 0.345 & $0.0$ & 0.475 & $0.0$ \\
AdaptGCGRU      & $-$ DY prior                          & 0.345 & $-0.0$ & 0.510 & $+7.5$ \\
GCGRU\_dy      & $-$ learned attention                 & 0.343 & $-0.7$ & 0.486 & $+2.3$ \\
GCGRU\_corr    & prior $\to$ correlation graph         & 0.349 & $+0.9$ & 0.513 & $+8.1$ \\
GCGRU\_geo     & prior $\to$ geographic graph          & 0.346 & $+0.1$ & 0.486 & $+2.4$ \\
LSTM            & $-$ graph (temporal module only)      & 0.348 & $+0.8$ & 0.603 & $+27.0$ \\
HARX            & $-$ neural module                     & 0.364 & $+5.4$ & 0.667 & $+40.6$ \\
HAR             & $-$ global factor                     & 0.396 & $+14.6$ & 1.009 & $+112.7$ \\
\bottomrule
\end{tabular}
\caption{Component analysis. Each row removes or replaces one
component of the full model; $\Delta$ is the change in pooled
(respectively crisis-day) QLIKE relative to the full model, positive
meaning worse. All rows are estimated under the identical
expanding-window protocol; five-seed dispersion and Diebold-Mariano
tests against the full model accompany the replication table in the
repository.}
\label{tab:ablation}
\end{table}
\noindent The pooled results show that all graph-based models perform similarly. Removing the learned attention while keeping the fixed DY graph slightly improves QLIKE by $0.7\%$, indicating that the DY network already captures most of the useful cross-market information. In contrast, the differences become clear during crisis periods. Removing the DY prior increases crisis QLIKE by $7.5\%$, replacing it with the correlation graph increases it by $8.1\%$, and removing the graph entirely (LSTM) increases it by $27.0\%$. Simpler models perform even worse, with HARX and HAR increasing crisis QLIKE by $40.6\%$ and $112.7\%$, respectively. These results show that the DY network is the most important component during crises, while its contribution is much smaller during calm periods.


\subsection{Robustness Across Subsamples} 
To examine whether the results are driven by the COVID-19 market crash, we repeat the analysis after removing $2020$ from the out-of-sample period. Figure~\ref{fig:yearly} (\ref{appendix:B}) shows that a graph-based model achieves the lowest annual QLIKE in four of the six years ($2020$, $2022$, $2023$, and $2024$), while the network-blind LSTM performs best in $2021$ and GBM in $2025$. In all years, the graph-based models remain close to the best performer.
\begin{table}[H]
\centering\small
\begin{tabular}{lcccc}
\toprule
 & \multicolumn{2}{c}{Crisis QLIKE} & \multicolumn{2}{c}{DM vs LSTM (crisis), $p$} \\
\cmidrule(lr){2-3}\cmidrule(lr){4-5}
Model & from 2020 (71 d) & from 2021 (42 d) & from 2020 & from 2021 \\
\midrule
HAR            & 1.009 & 0.918 & - & - \\
LHAR           & 0.586 & 0.629 & - & - \\
GBM            & 0.682 & 0.568 & - & - \\
LSTM           & 0.603 & 0.589 & - & - \\
GCGRU\_dy     & 0.486 & 0.534 & 0.023 & 0.374 \\
AdaptGCGRU\_dy & 0.475 & 0.509 & 0.018 & 0.235 \\
\bottomrule
\end{tabular}
\caption{Crisis-day performance with and without 2020. Crisis days are
the $[0,+20]$ windows after the out-of-sample events; the from-2021
columns remove COVID and retain the Ukraine-invasion and tariff-shock
windows. DM tests the daily crisis-day loss differential of each graph
model against the network-blind LSTM.}
\label{tab:sens}
\end{table}
\noindent Table~\ref{tab:sens} reports the crisis-period results after excluding $2020$, leaving only the $42$ crisis days associated with the Ukraine invasion and the $2025$ tariff shock. The main conclusions remain unchanged. The graph models continue to achieve the lowest crisis QLIKE, with the adaptive and fixed-DY models outperforming the network-blind LSTM by $13.7\%$ and $9.4\%$, respectively. However, the DM test is no longer statistically significant because only $42$ crisis days remain. This indicates that the graph models are robust across subsamples, while the stronger statistical evidence in the full sample is mainly due to the larger number of COVID crisis observations.


\subsection{Robustness to the Volatility Proxy} \ref{Appendix:A2} discusses the limitation of the Garman--Klass (GK) estimator, which does not include overnight returns. To examine whether our results depend on this choice, we re-evaluate all models using the squared close-to-close return as an alternative volatility proxy \cite{patton2011}. The main conclusions remain unchanged. The same model ranks first, the top four models are unchanged, and all graph-based models continue to outperform HAR significantly ($p<0.001$). Only the ordering of the classical benchmark models changes slightly. These results show that our conclusions are robust to the choice of volatility proxy.

\subsubsection{Robustness to Forecast Frequency}

To examine whether our results are influenced by differences in market trading hours, we repeat the analysis using weekly data ($574$ weeks), where such timing differences are much smaller. The main conclusions remain unchanged. Both the graph model and the network-blind LSTM continue to outperform HAR, with QLIKE values of $0.2325$ and $0.2321$, respectively. However, the graph model and the network-blind LSTM perform almost identically ($0.2325$ vs.\ $0.2321$). This suggests that the benefit of the graph comes mainly from capturing cross-market volatility spillovers at the daily frequency. By the weekly horizon, most of these spillovers have already occurred, leaving little additional information for the network to exploit.\\\\
In addition, the results are stable across different random initializations and refitting periods. For all graph-based models, the worst-seed DM statistic remains above 3.7, and no single six-month refit period accounts for the observed performance gains. This indicates that the results are robust to both random initialization and the choice of estimation period.

\subsection{Interpretation} Three findings deserve economic interpretation. First, the larger forecasting gains during crisis periods are an expected feature of the proposed approach. During calm markets, volatility changes slowly and is mainly driven by local market conditions, so a univariate model performs well. Consequently, the network-based and network-blind models show similar performance because cross-market spillovers are limited. When a major market shock spreads across countries within a day or two, the network structure helps improve forecasting accuracy. Accurate volatility forecasts are especially important during such periods for risk managers and clearinghouses, where timely risk assessment is critical.\\

\noindent Second, the data-inferred network closely matches the DY network, even though the two are constructed using very different methods. Both identify the US markets as the main transmitters of shocks and the European markets as a closely connected group. This suggests that the observed spillover structure is a stable feature of global equity markets rather than an artifact of a particular estimation method.\\\\
\noindent Third, the statistical and economic results are consistent. The models that achieve the lowest QLIKE also provide the highest economic value in the volatility-targeting strategy. This shows that the improved forecasting accuracy translates into better investment performance and is not simply the result of biased or conservative forecasts. Therefore, both the statistical and economic results support the effectiveness of the proposed models.

\section{Conclusion}\label{section:7}
Conditioning each market's volatility forecast on the DY spillover network improves one-day-ahead realized volatility forecasts for $29$ global equity indices by $13\%$ in QLIKE compared with HAR. The improvement is supported by the Diebold--Mariano tests, the model confidence set, and robustness checks. In a volatility-targeting strategy, the DY-based model delivers more than $320$ bp of annual certainty-equivalent gain after transaction costs and is $21\%$ more accurate than the network-blind model during crisis periods.

\noindent Among the network structures considered, the DY network performs best. It achieves the highest model-confidence-set $p$-value, while HAR, LHAR, GBM, the correlation graph, and the random walk are excluded from the model confidence set. The data-inferred network also closely matches the DY network, supporting its use as an informative prior for forecasting global market volatility.
The results suggest that graph choice becomes particularly important during periods of elevated cross-market stress, even when differences in pooled forecast accuracy are relatively modest. Incorporating economically meaningful network structures can substantially improve graph-based financial forecasting.

\noindent The study has some limitations. The Garman-Klass volatility proxy excludes overnight returns, the DY network is updated semi-annually rather than daily, and only equity markets are considered. After removing the COVID period, only $42$ crisis days remain, reducing the statistical power of the crisis analysis, although the graph models continue to outperform the network-blind model. Future work will consider dynamically updated networks, intraday volatility measures, additional asset classes, and multi-step forecasting.

\section*{Data and Code Availability}
The code and data used in this study are available at the following GitHub repository: \url{https://github.com/nishit-soni/volatility-spillover-forecasting}.

\bibliographystyle{abbrv}
\bibliography{references}
\newpage
\appendix
\renewcommand{\thesection}{Appendix \Alph{section}}
\section{Data Construction and Cleaning}\label{appendix:A}

This appendix provides additional details related to the data construction described in Section~\ref{section:3}. Specifically, it covers the union trading calendar and coverage screens (\ref{Appendix:A1}), the properties of the Garman-Klass range estimator (\ref{Appendix:A2}), the despiking, winsorization, and missing-value rules (\ref{Appendix:A3}), and two vendor-specific notes (\ref{Appendix:A4}).


\subsection{The Union Trading Calendar and Sample Coverage}\label{Appendix:A1}
The 29 markets follow different national holiday schedules, so no single
exchange calendar covers all of them. We therefore work on a
\emph{union calendar}: the set of all dates on which at least one of the
29 indices traded, comprising 2{,}866 trading days between 2 January
2015 and 30 December 2025. A market enters the study only if it has a
native observation on at least 50\% of these union-calendar days and a
live price feed through the end of the sample; both screens are applied
ex ante, so the universe is fixed before any forecast is formed and is
free of survivorship-related look-ahead. On a day when a given market's
exchange is closed while at least one other market trades, that market
simply has no native observation for the date; these non-synchronous
holidays, rather than any data error, are the dominant source of the
missing values discussed in \ref{Appendix:A3}.

\subsection{The Garman-Klass Range Estimator}\label{Appendix:A2}
Because intraday transaction data are not available for a panel this
broad, each day's variance is proxied by the range-based estimator of
\cite{garman1980} in Equation~\eqref{eq:gk}, which combines the squared
log high-low range with the squared log close-open return. The
high-low range uses the full intraday price path rather than two
endpoints and is consequently a much more informative signal of the
day's variance than the squared close-to-close return; under the
estimator's maintained assumptions of zero drift and continuous
within-session trading, it is roughly five to eight times more
efficient, the property that makes a daily international
realized-variance panel tractable. Two assumptions qualify the measure.
First, it uses only within-session prices and therefore omits the
overnight return between one day's close and the next day's open, so
markets with large or systematic overnight gaps carry modestly
understated variance \emph{levels}. Because this is a roughly constant
level shift within each market and the QLIKE loss is computed market by
market, it affects estimated levels but not the relative ranking of
competing forecasts. Second, the recorded high and low are discrete
extremes of a path that is only sampled, not observed continuously,
which biases the range estimator slightly downward, again a level
effect rather than a source of spurious cross-market predictability.

\subsection{Despiking, Winsorization, and Missing-Value Treatment}\label{Appendix:A3}
Three cleaning steps are applied to the raw series. (i)~\emph{Despiking}:
single-day absolute returns above 25\% are treated as vendor errors
rather than genuine moves and set to missing; the sample contains one
such observation for Argentina's MERVAL and two for South Africa's JSE.
(ii)~\emph{Winsorization}: as a further guard against fat-tailed vendor
noise, the return series of these two markets are winsorized at their
1st and 99th percentiles. (iii)~\emph{Degenerate ranges}: 39 market-days
on which the recorded high equals the low produce a zero range and an
undefined log-variance, and are set to missing; these are concentrated
in holiday-adjacent sessions. After these steps, $5.70\%$ of the
realized-variance panel is missing, almost entirely because of the
non-synchronous holidays described in \ref{Appendix:A1}. For model inputs
only, missing values are filled with the trailing median of the
market's realized variance, which uses no future information; filled
values are never included in any loss calculation, so no forecast is
ever scored against an imputed target.

\subsection{Vendor-Specific Notes}\label{Appendix:A4}
Two markets have idiosyncratic feed characteristics that are retained
rather than corrected. Israel's TA125 enters Yahoo Finance's history
only in 2017, so its pre-2017 observations are treated as missing rather
than back-filled. Saudi Arabia's TASI trades on a Sunday-to-Thursday
week; its observations are placed on the union calendar on their true
trading dates, and the resulting Friday gaps are treated as missing.
Both indices are kept because they supply genuine Middle-Eastern breadth
that no other market in the universe provides. Finally, as noted in \ref{Appendix:A2}, the Garman-Klass estimator omits overnight returns, and
net-spillover directionality is additionally sensitive to the order in
which time zones close; both are therefore interpreted qualitatively
throughout the paper.

\section{Additional Diagnostics of the Learned Network}\label{appendix:B}

This appendix collects supplementary views of the learned attention
network and the robustness analyses of Section~\ref{section:6}. All
heatmaps use a common perceptually-uniform colour scale with markets
grouped into regional blocks.

\begin{figure}[htbp]
\centering
\includegraphics[width=0.92\textwidth]{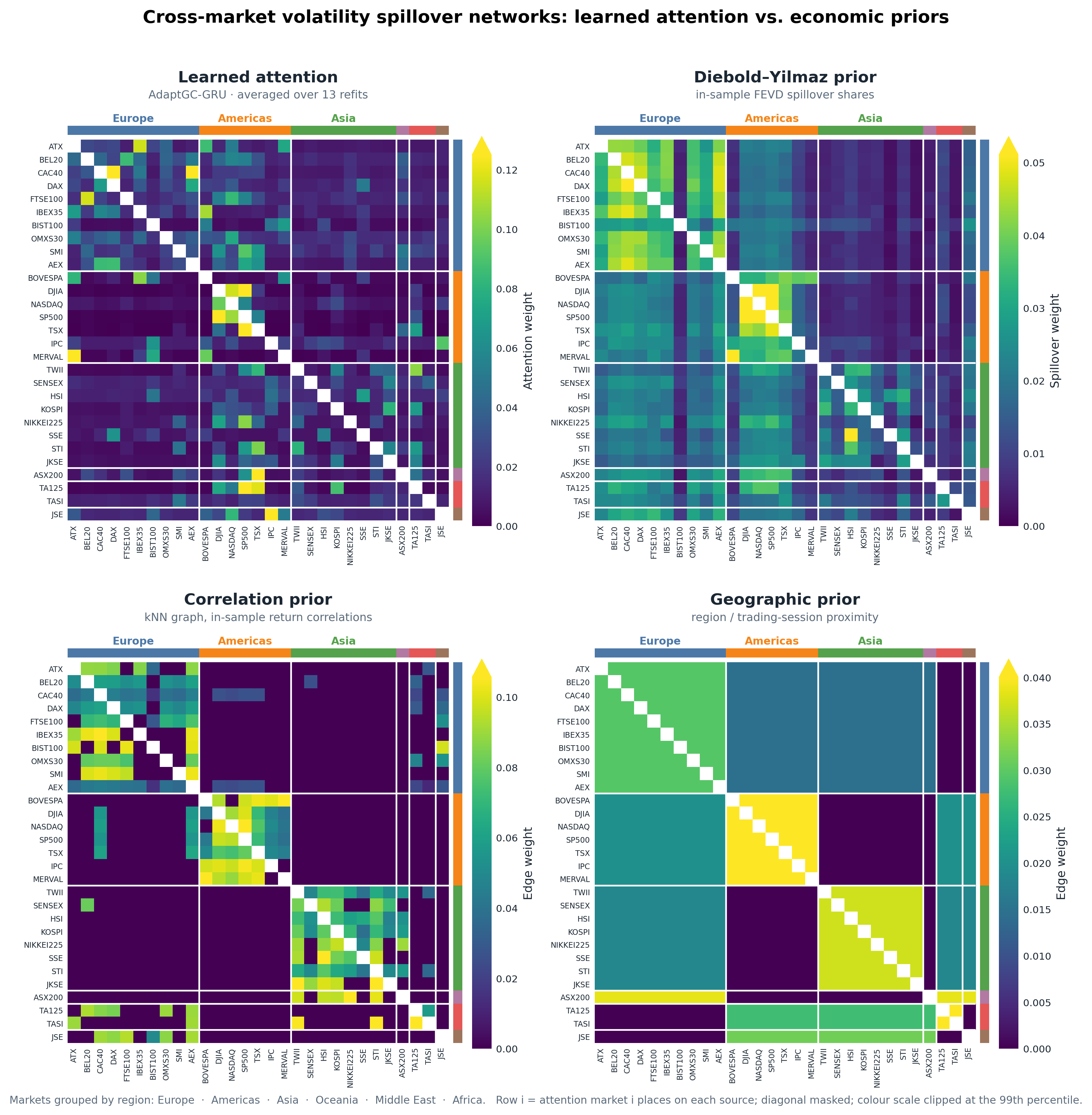}
\caption{Learned attention versus all three priors (Diebold-Yilmaz,
correlation-$k$NN, geographic).}
\label{fig:adjall}
\end{figure}

\begin{figure}[p]
\centering
\includegraphics[width=0.8\textwidth]{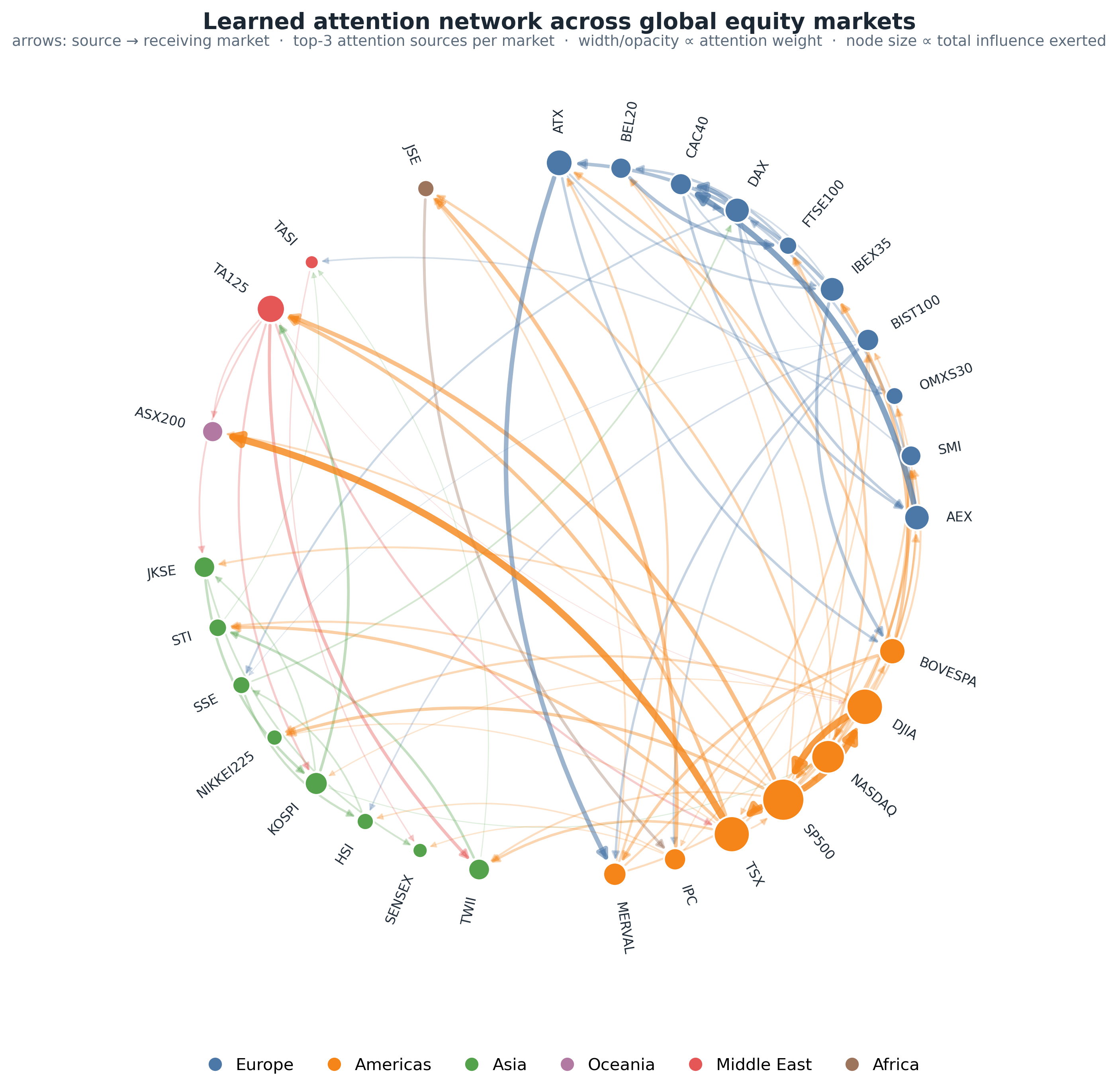}
\caption{The learned attention network: top-3 sources per market;
node size proportional to influence exerted; colours denote regions.}
\label{fig:attgraph}
\end{figure}

\begin{figure}[p]
\centering
\includegraphics[width=0.95\textwidth]{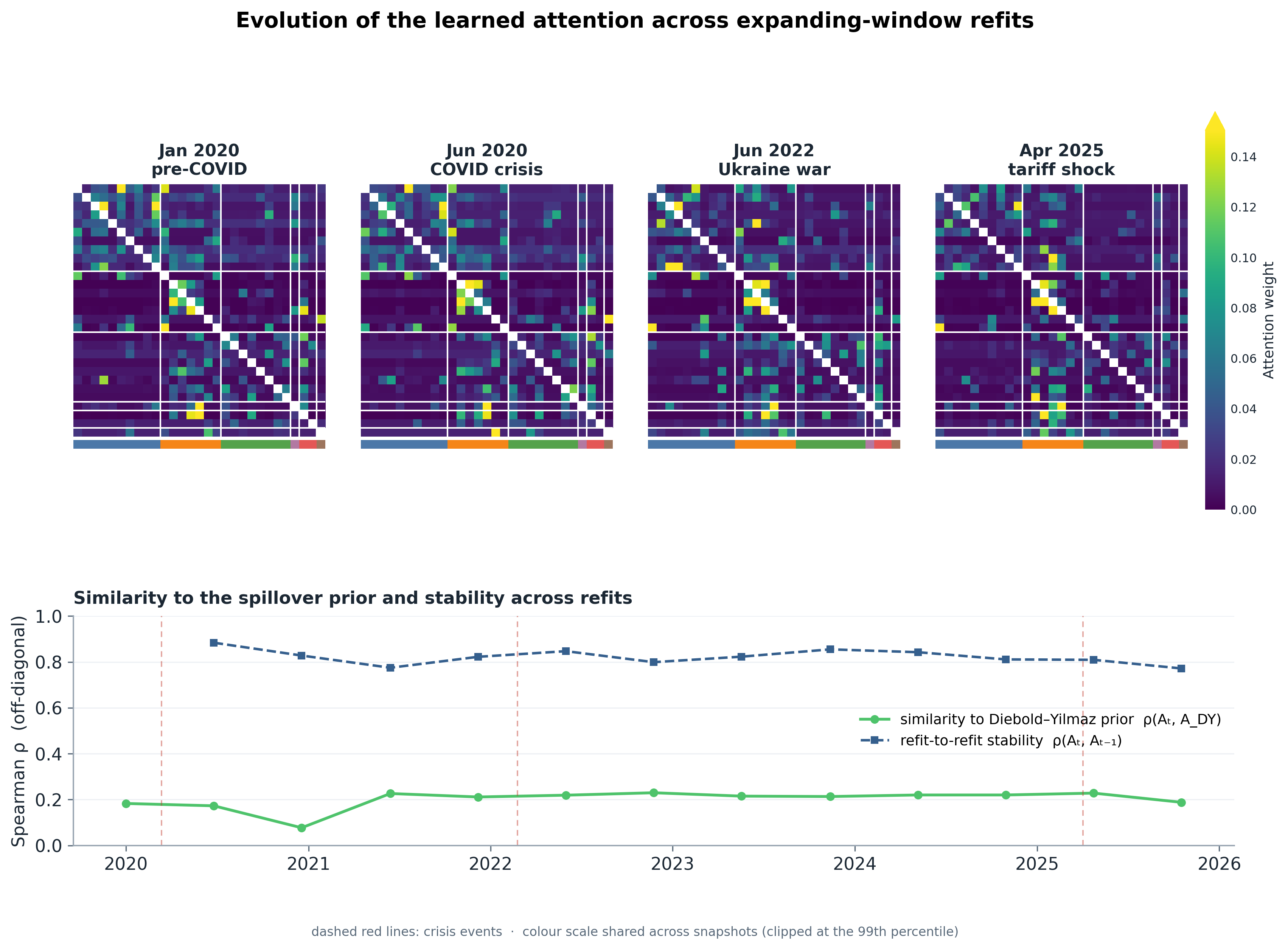}
\caption{The learned attention at four refits (pre-COVID, COVID,
Ukraine war, 2025 tariff shock; shared colour scale), and its
similarity to the DY prior and refit-to-refit stability across all 13
refits.}
\label{fig:attdyn}
\end{figure}

\begin{figure}[p]
\centering
\includegraphics[width=0.78\textwidth]{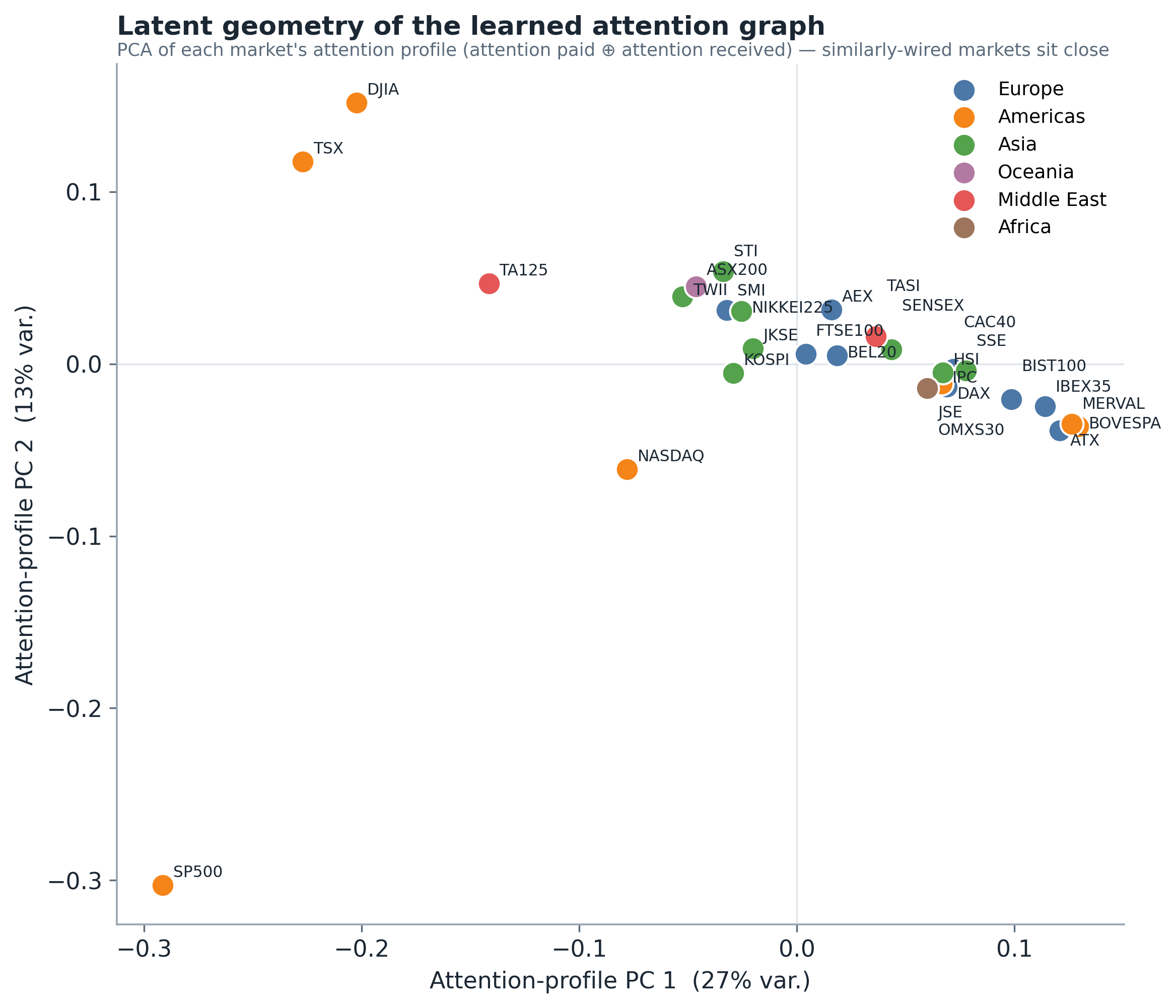}
\caption{Principal-component projection of each market's attention
profile. Similarly-wired markets sit close together.}
\label{fig:attembed}
\end{figure}

\begin{figure}[p]
\centering
\includegraphics[width=\textwidth]{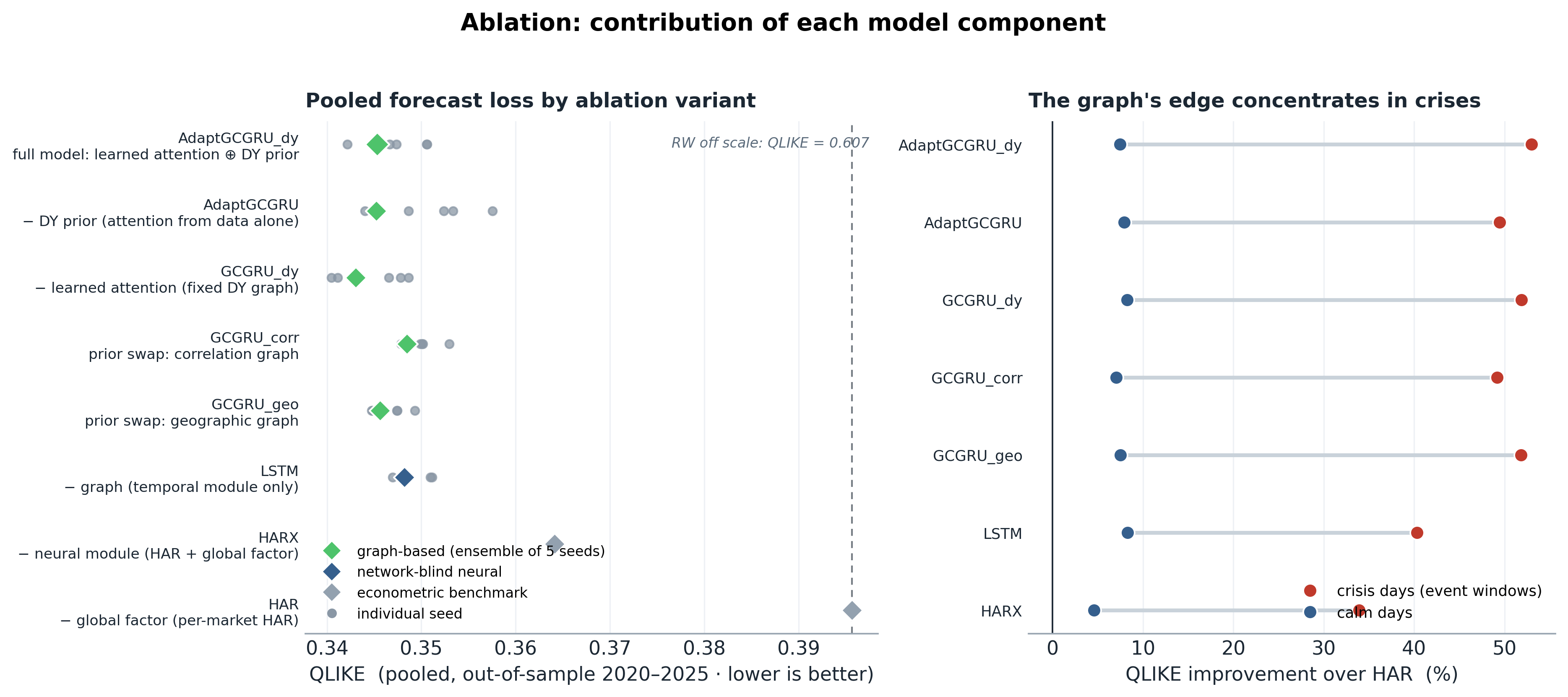}
\caption{The component ablation of Table~\ref{tab:ablation}: pooled
loss per variant with per-seed dispersion (left) and the crisis versus
calm improvement over HAR (right).}
\label{fig:ablfig}
\end{figure}

\begin{figure}[p]
\centering
\includegraphics[width=0.9\textwidth]{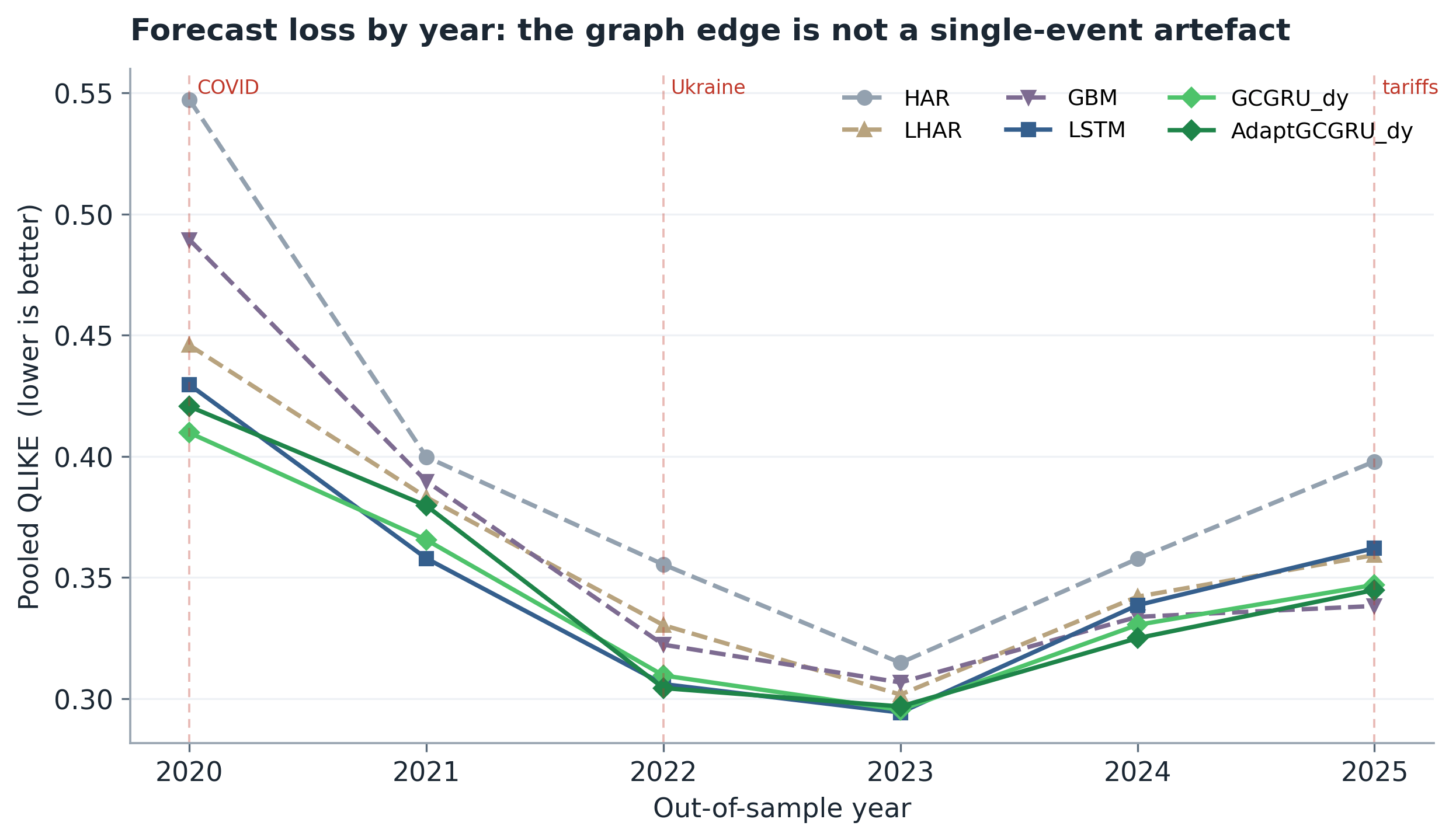}
\caption{Pooled QLIKE by out-of-sample year for the key models.}
\label{fig:yearly}
\end{figure}

Figure~\ref{fig:adjall} adds the correlation-$k$NN prior to the
comparison of Figure~\ref{fig:adj}. Figure~\ref{fig:attgraph} draws
the learned network directly: each market's three strongest attention
sources, with edge width proportional to the weight, make the North
American hub and the intra-European block visible at a glance.
Figure~\ref{fig:attdyn} tracks the learned network across the 13
semi-annual refits: it is highly stable from refit to refit (Spearman
$\rho$ between consecutive refits of $0.78$-$0.88$) while its
rank correlation with the DY prior fluctuates in the $0.08$-$0.23$
range, confirming that the recovery of the DY structure documented in
Section~5.3 is a persistent property of the estimates rather than an
artifact of a single refit. Because averaging across the 13 refits removes refit-specific noise, the refit-averaged network used in Section~5.3 correlates more strongly with the DY prior ($\rho = 0.254$) than any single refit
does (0.08 to 0.23). Figure~\ref{fig:attembed} projects each
market's attention profile (attention paid and received) onto its two
principal components ($27\%$ and $13\%$ of variance): SP500 sits alone
as the system's dominant hub, with the remaining North American
indices adjacent and the regional blocks otherwise preserved.
Figure~\ref{fig:ablfig} visualizes the component ablation of
Table~\ref{tab:ablation}, and Figure~\ref{fig:yearly} the by-year
losses underlying the subsample analysis.

\end{document}